 \documentclass[final,3p,times,twocolumn]{elsarticle}

\usepackage{amssymb}

\journal{Astroparticle Physics}

\begin{document}

\begin{frontmatter}



\title{Potential for Precision Measurement of Low-Energy Antiprotons with GAPS for Dark Matter and Primordial Black Hole Physics}


\author[CAL]{T. Aramaki} 
\ead{tsuguo@astro.columbia.edu}

\author[UCB]{S. E. Boggs} 
\author[Hawaii]{P. von Doetinchem}
\author[JAXA]{H. Fuke} 
\author[CAL]{C. J. Hailey}
\author[UCLA]{S.A.I. Mognet}
\author[UCLA]{R. A. Ong}
\author[CAL]{K. Perez}
\author[UCLA]{J. Zweerink} 

\address[CAL]{Columbia Astrophysics Laboratory, Columbia University, New York, NY 10027, USA}
\address[UCB]{Space Sciences Laboratory, University of California, Berkeley, CA 94720, USA}
\address[Hawaii]{Department of Physics and Astronomy, University of Hawaii, Honolulu, HI 96822, USA}
\address[JAXA]{Institute of Space and Astronautical Science, Japan Aerospace Exploration Agency (ISAS/JAXA), Sagamihara, Kanagawa 229-8510, Japan}
\address[UCLA]{Department of Physics and Astronomy, University of California, Los Angeles, CA 90095, USA}

\begin{abstract}

The general antiparticle spectrometer (GAPS) experiment is a proposed indirect dark matter search focusing on antiparticles produced by WIMP (weakly interacting massive particle) annihilation and decay in the Galactic halo. In addition to the very powerful search channel provided by antideuterons \cite{Donato2000,Donato2008,Vittino2013,Fornengo2013a}, GAPS has a strong capability to measure low-energy antiprotons ($0.07 \le E \le$ 0.25 GeV) as dark matter signatures. This is an especially effective means for probing light dark matter, whose existence has been hinted at in the direct dark matter searches, including the recent result from the CDMS-II experiment \cite{Agnese2013}. While severely constrained by LUX and other direct dark matter searches \cite{Akerib2013}, light dark matter candidates are still viable in an isospin-violating dark matter scenario and halo-independent analysis \cite{Nobile2013a,Nobile2013b}. Along with the excellent antideuteron sensitivity, GAPS will be able to detect an order of magnitude more low-energy antiprotons, compared to BESS \cite{Abe2012,Orito2000}, PAMELA \cite{Adriani2010} and AMS-02 \cite{Casaus2009}, providing a precision measurement of low-energy antiproton flux and a unique channel for probing light dark matter models. Additionally, dark matter signatures from gravitinos and Kaluza-Klein right-handed neutrinos as well as evidence of primordial black hole evaporation can be observed through low-energy antiproton search.

\end{abstract}


\end{frontmatter}



\section{Introduction}
\label{Sec:Introduction}

Although the recent result by the Planck experiment shows that 27\% of our universe is composed of dark matter \cite{Planck2013}, the nature and origin of dark matter are still unknown. This is one of the greatest cosmological puzzles of the 21st century. WIMPs are very well-motivated candidates for dark matter. Neutralinos in supersymmetric (SUSY) theories and right-handed neutrinos in extra dimension theories are examples of WIMP candidates \cite{Cirelli2012,Lavalle2012}. 

In the past decade, the DAMA/LIBRA \cite{Bernabei2010}, CoGeNT \cite{Aalseth2011a,Aalseth2011b,Aalseth2011c}, and CRESST-II \cite{Angloher2011} experiments claimed signals from light dark matter ($m < $ 30 GeV/c$^2$) and CDMS-II recently reported that they found three WIMP-candidate events with a dark matter mass of $\sim$ 10 GeV/c$^2$ \cite{Agnese2013}. Although this model has been/will be further probed by LUX and other dark matter experiments \cite{Akerib2013}, light dark matter candidates remain viable in an isospin-violating scenario and halo-independent analysis, as discussed in \cite{Nobile2013a,Nobile2013b}. While dark matter couplings to protons and neutrons can generally enhance the interaction cross-section between dark matter and the target atom, it can be reduced in the isospin-violating scenario, and thus the dark matter interaction can be greatly suppressed in LUX and other experiments. Moreover, as the nuclear recoil energy in a direct search is strongly affected by the dark matter mass and the velocity of the dark matter halo, the recoil energy from the light dark matter may not be large enough in comparison to the low-energy threshold of the experiments. Background study and discrimination are the key for direct detection of light dark matter candidates, due to high background rates and electronics noise near the energy threshold. 


The GAPS experiment, described in the following sections, has the capability to search for light dark matter. While being complementary to direct and other indirect dark matter searches as well as collider experiments, GAPS' low-energy antiproton measurement, along with its antideuteron sensitivity, will be able to probe light dark matter models with completely different detection methods \cite{Hailey2009,Aramaki2014}. 

In comparison to BESS \cite{Abe2012,Orito2000}, PAMELA \cite{Adriani2010} and AMS-02 \cite{Casaus2009}, carrying out the new GAPS experiment will not only open new phase space for measuring antiprotons at low energy (kinetic energy: $E < 0.25$ GeV), but will also offer important complementarity in terms of identification technique \cite{Aramaki2014}. The BESS and AMS-02 antiparticle measurements face similar backgrounds as they both strongly rely on the track curvature measurement in their magnetic fields, while GAPS provides an independent antiproton measurement by utilizing exotic atom capture and decay. Furthermore, the foreseen GAPS flight trajectory from Antarctica is favorable for low-energy cosmic-ray measurements because of the low geomagnetic cutoff in comparison to the flight trajectory of AMS-02 on the International Space Station at rather high geomagnetic cutoffs. 

In this paper, an overview of the GAPS project will be given in Section \ref{Sec:GAPS} and the GAPS antiproton search for dark matter and evaporating primordial black holes (PBHs) will be discussed in Section \ref{Sec:Antiproton}.  

\section{GAPS project}
\label{Sec:GAPS}

\begin{figure}[!b]
\begin{center} 
\includegraphics*[width=7.5cm]{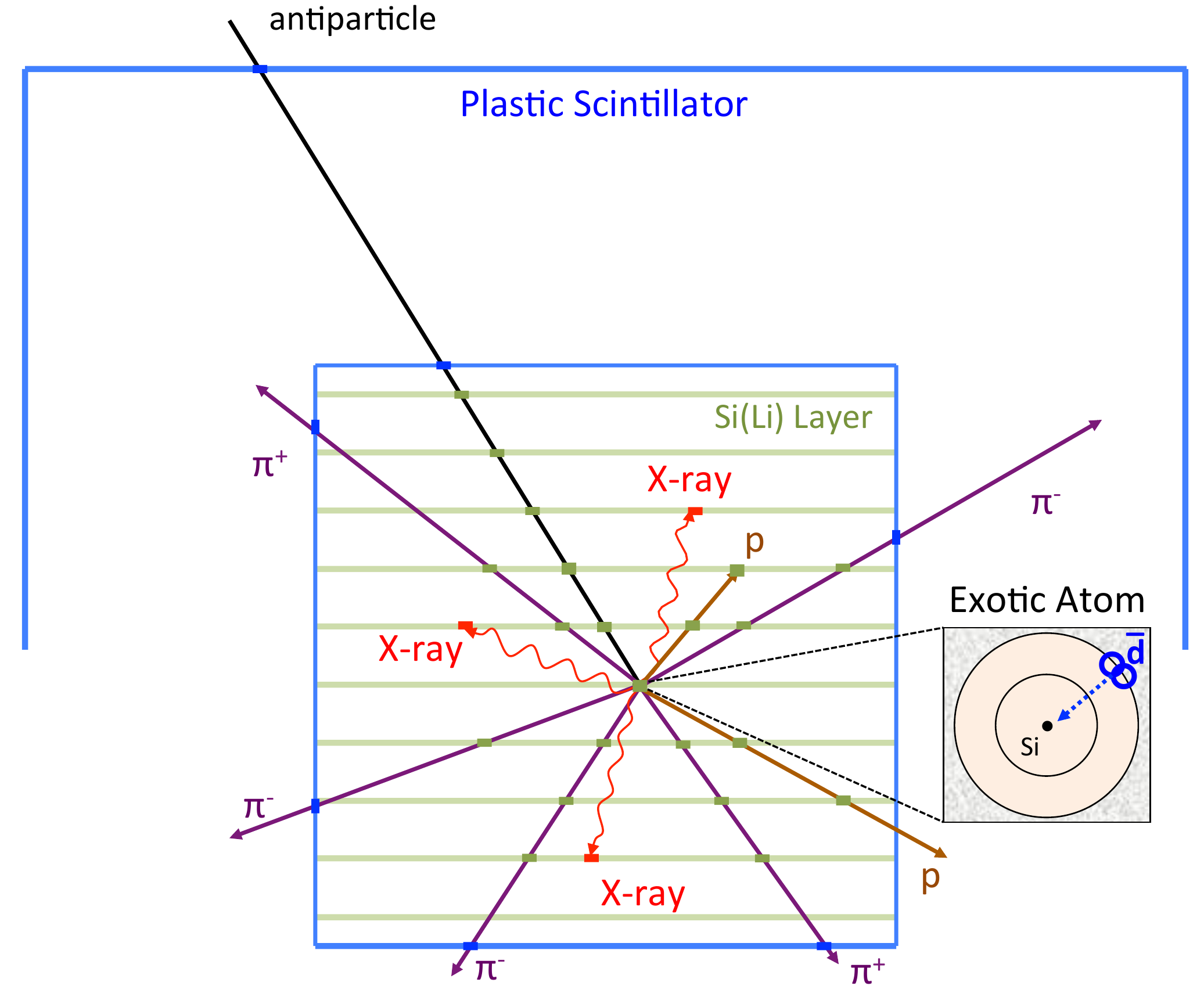}
\end{center}
\caption{Schematic view of GAPS instrument and detection method. 10 layers of Si(Li) detectors are surrounded by the TOF plastic scintillators. An antiparticle slows down and stops in the Si(Li) target forming an exotic atom. The atomic X-rays are emitted as it de-excites, followed by charged particle emission from the nuclear annihilation.}
\label{detection}
\end{figure} 

The GAPS project is a balloon-based indirect dark matter search that is optimized for low-energy antiparticles produced by dark matter annihilation and decay in the Galactic halo \cite{Mori2002,Hailey2004}. The GAPS detection method involves capturing antiparticles in a target material with the subsequent formation and decay of an excited exotic atom \cite{Hailey2009,Aramaki2013}. A time-of-flight (TOF) system measures the velocity (energy) of an incoming antiparticle. The particle slows down in the instrument by dE/dX energy loss and stops in the target material, forming an excited exotic atom. The exotic atom de-excites with the emission of Auger electrons as well as atomic X-rays. Given the atomic number of the target and the
Bohr formula for the atomic X-ray energy, the measured X-ray energies uniquely determine the mass of the captured antiparticle. Eventually, the antiparticle is captured by the nucleus in the atom, where it annihilates with the emission of pions and protons. The number of annihilation products provides an additional identification for the captured antiparticle. This process is illustrated in Figure \ref{detection}. Note that some antiparticles ($E >$ 0.15 GeV/n) annihilate before stopping (inflight annihilation), but they can be identified with the particle multiplicity of the annihilation products.  

The GAPS instrument has 10 layers of lithium-drifted silicon (Si(Li)) detectors surrounded by TOF plastic scintillators (see Figure \ref{detection}). Each layer is composed of 4 inch diameter, 2.5 mm thick Si(Li) detectors, segmented into four channels \cite{Aramaki2012,Perez2013}. This provides modest three-dimensional particle tracking, allowing us to count the number of particles produced in the nuclear annihilation and separately identify atomic X-rays from particle tracks. 
 
The GAPS detection method was evaluated in an antiproton beam test in 2005 \cite{Aramaki2013,Hailey2006} and a prototype flight (pGAPS) was successfully conducted in 2012 \cite{Doetinchem2014,Mognet2014,Fuke2013}. A first GAPS science flight is proposed from Antarctica in the austral summer of 2017-2018 (solar minimum periods).

\section{GAPS antiproton search}
\label{Sec:Antiproton}

\subsection{Overview}
Annihilation and decay of dark matter can produce copious antiprotons. Although it is challenging to distinguish the primary antiproton flux (due to dark matter annihilation and decay) from the secondary flux (due to cosmic-ray interactions with the interstellar medium), a dark matter signature could be seen as an excess in the antiproton flux at low energy above the predicted astrophysical flux. Since the secondary antiproton flux steeply decreases at low energy (see Figure \ref{BG}), the spectrum shape would become flat or show a bump if a primary component existed. 

The antiproton search by BESS-Polar II \cite{Abe2012} and PAMELA \cite{Adriani2010} reported no excess in the antiproton flux and provided constraints on the parameter space of dark matter mass and annihilation cross-section for dark matter models \cite{Cirelli2013,Fornengo2013b}. The diffuse gamma-ray emission measured by Fermi-LAT also provides constraints on the same parameter space \cite{Ackermann2013,Ackermann2011}. 

Light dark matter models, however, are not completely ruled out, since the primary antiproton flux significantly varies based on astrophysical parameters, such as the propagation model, the Galactic halo density profile and the solar modulation \cite{Cirelli2013,Ackermann2013,Ackermann2011}. The dark matter annihilation or decay takes place in the dark matter halo, while the secondary antiprotons can be produced only in the Galactic disk, and thus the constraints obtained by the boron to carbon ratio induce a much larger uncertainty on the primary antiproton flux than on the secondary flux \cite{Donato2000,Salati2010}. The primary flux will also be reduced if the annihilation cross-section is velocity suppressed \cite{Boehm2013} or if there is more than one type of dark matter \cite{Salati2010,Bottino2003}, which decreases the dark matter annihilation rate and the antiproton production rate in the Galactic halo.  

Moreover, the low-energy antiproton data are very limited and it is difficult to constrain models of the primary and secondary antiproton fluxes. BESS (PAMELA) detected only $\sim$ 30 (10) antiprotons at $E \sim$ 0.2 GeV, and no events were detected at $E \le$ 0.1 GeV \cite{Abe2012,Orito2000,Adriani2010}. On the other hand, GAPS will be able to detect more than $\sim 10^3$ antiprotons at $0.07 \le E \le 0.25$ GeV in one LDB (long duration balloon) flight ($\sim$ 40 days) with approximately seven data points in this energy range ($\sim 200$ counts each) \cite{Aramaki2014}, which will allow us to precisely determine the spectral shape of the antiproton flux at low energy. Therefore, GAPS provides a precision low-energy antiproton flux measurement and has a strong potential to observe dark matter signatures.  

\subsection{Antiproton identification}

As discussed in the previous section, GAPS utilizes atomic X-rays and annihilation products from the decay of exotic atoms in order to identify antiparticles. Unlike antideuteron sensitivity, where the main background is antiprotons, antiprotons can be easily identified from other particles since the main background is protons. Since slow protons ($\beta < 0.5$) triggered by the TOF system will not be able to form exotic atoms nor produce any relativistic secondary particles, the multiple simultaneous detection of relativistic pions provides a clear identification of antiprotons. We also expect to see simultaneous atomic X-rays in antiproton events. Note that the probability for two or more cosmic-ray protons simultaneously entering into the detector volume and stopping in the same channel is negligible. 

\subsection{Antiprotons from dark matter and PBH}

A dark matter signature could be seen as an excess in the low-energy antiproton flux. It would be prominent if the dark matter mass is relatively small ($m_{DM} <$ 30 GeV/c$^2$), as suggested by some direct dark matter searches. Light dark matter models exist within the phenomenological (effective) Minimal Supersymmetric Standard Model, pMSSM \cite{Boehm2013,Arbey2012,Mahmoudi2012} (eMSSM \cite{Bottino2003,Bottino2010,Bottino2005}) with a non-universal gaugino scenario, where there are 19 (8) free parameters. We will examine expected antiproton fluxes for three dark matter models and evaporating black holes below.   

\begin{figure}[!h]
\begin{center} 
\includegraphics*[width=7.5cm]{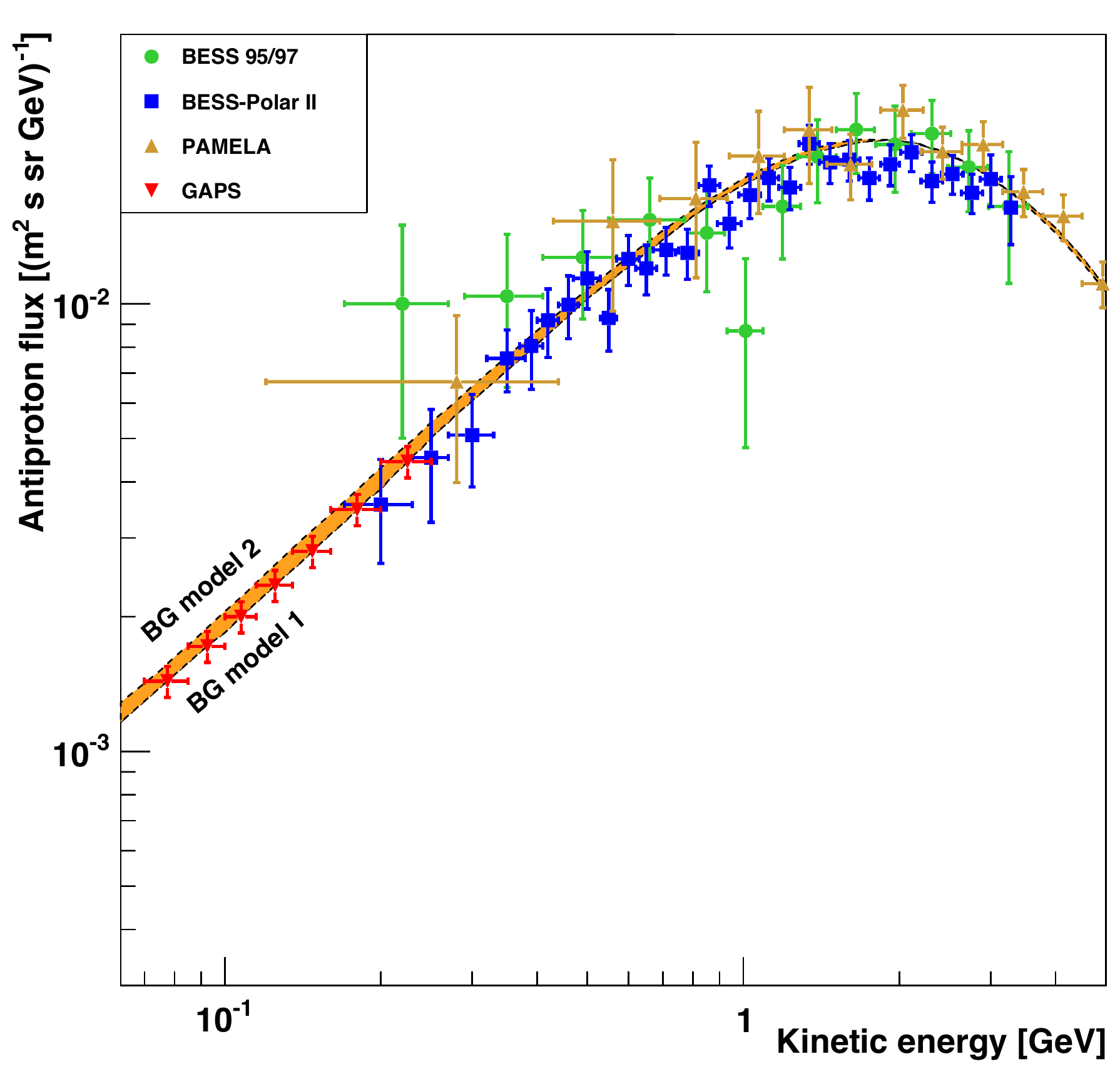}
\end{center}
\caption{Two different secondary antiproton flux models (black dashed lines, extrapolated down to $E$ = 65 MeV: BG model 1 for leaky box model with force-field modulation $\phi = 600$ MV \cite{Abe2012,Mitsui1996}, BG model 2 for leaky box drift model with negative solar magnetic field polarity and the tilt angle of the heliospheric current sheet = 15$^{\circ}$ \cite{Abe2012,Bieber1999}), together with the BESS-Polar II \cite{Abe2012}, BESS 95/97 \cite{Orito2000} and PAMELA \cite{Adriani2010} experimental data. The expected GAPS sensitivity for one LDB flight ($\sim$ 40 days, statistical uncertainty only) is also shown for the case of a pure secondary component (BG model 1).}
\label{BG}
\end{figure}

\begin{figure}[!h]
\begin{center} 
\includegraphics*[width=7.5cm]{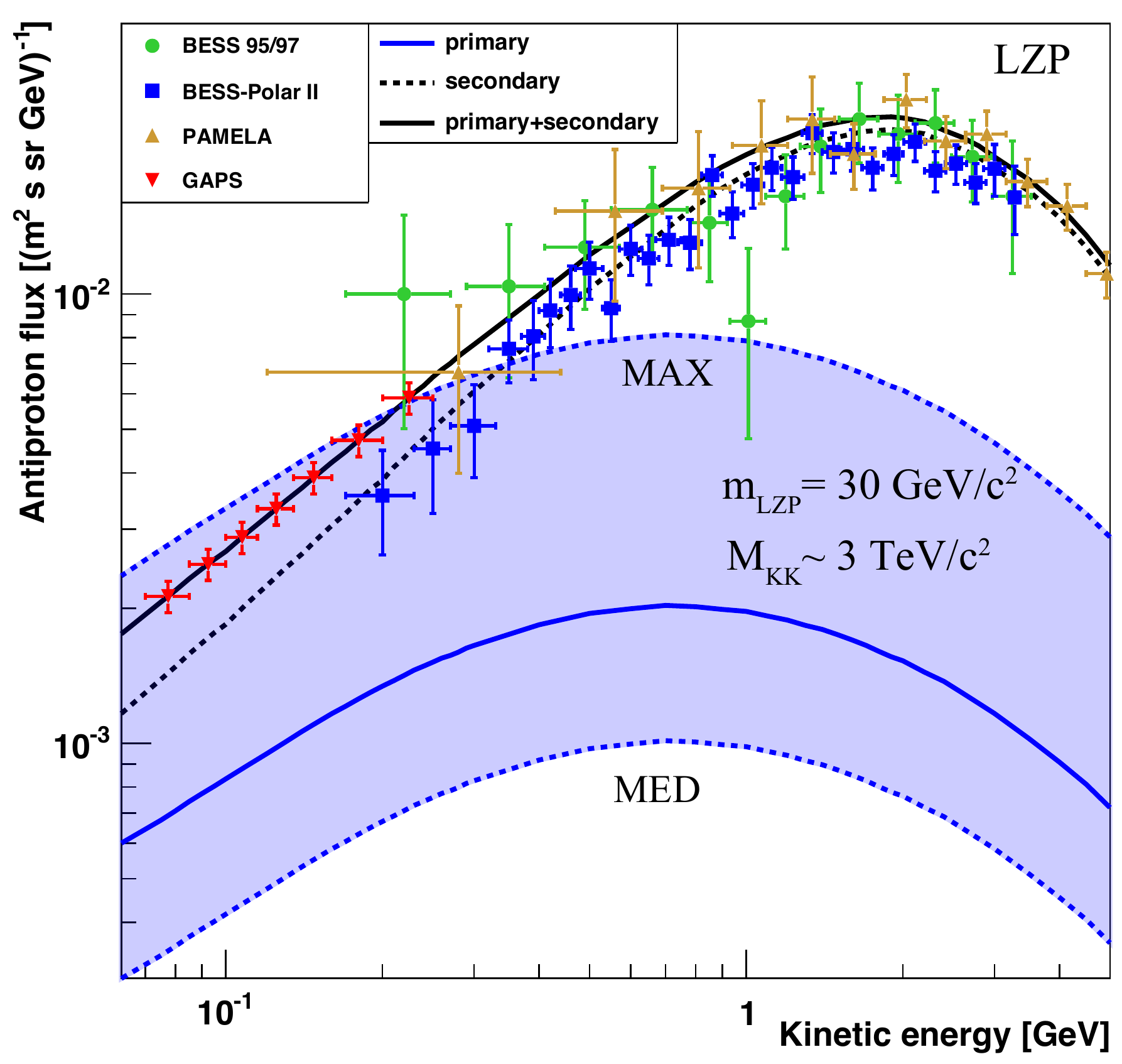}
\end{center}
\caption{The predicted primary antiproton fluxes at the top of the atmosphere (solar-minimum) from right-handed neutrino for $m_{\chi} \sim$ 30 GeV/c$^2$ with the medium and maximum astrophysical propagation models (blue dashed lines), together with the secondary antiproton flux (black dashed line \cite{Abe2012,Mitsui1996}) and the BESS-Polar II \cite{Abe2012}, BESS 95/97 \cite{Orito2000} and PAMELA \cite{Adriani2010} experimental data. The expected GAPS sensitivity for one LDB flight ($\sim$ 40 days, statistical uncertainty only) is also shown for the total antiproton flux (black solid line: primary flux (blue solid line) + secondary flux (black dashed line)).}
\label{KK}
\end{figure}

\begin{figure}[!h]
\begin{center} 
\includegraphics*[width=7.5cm]{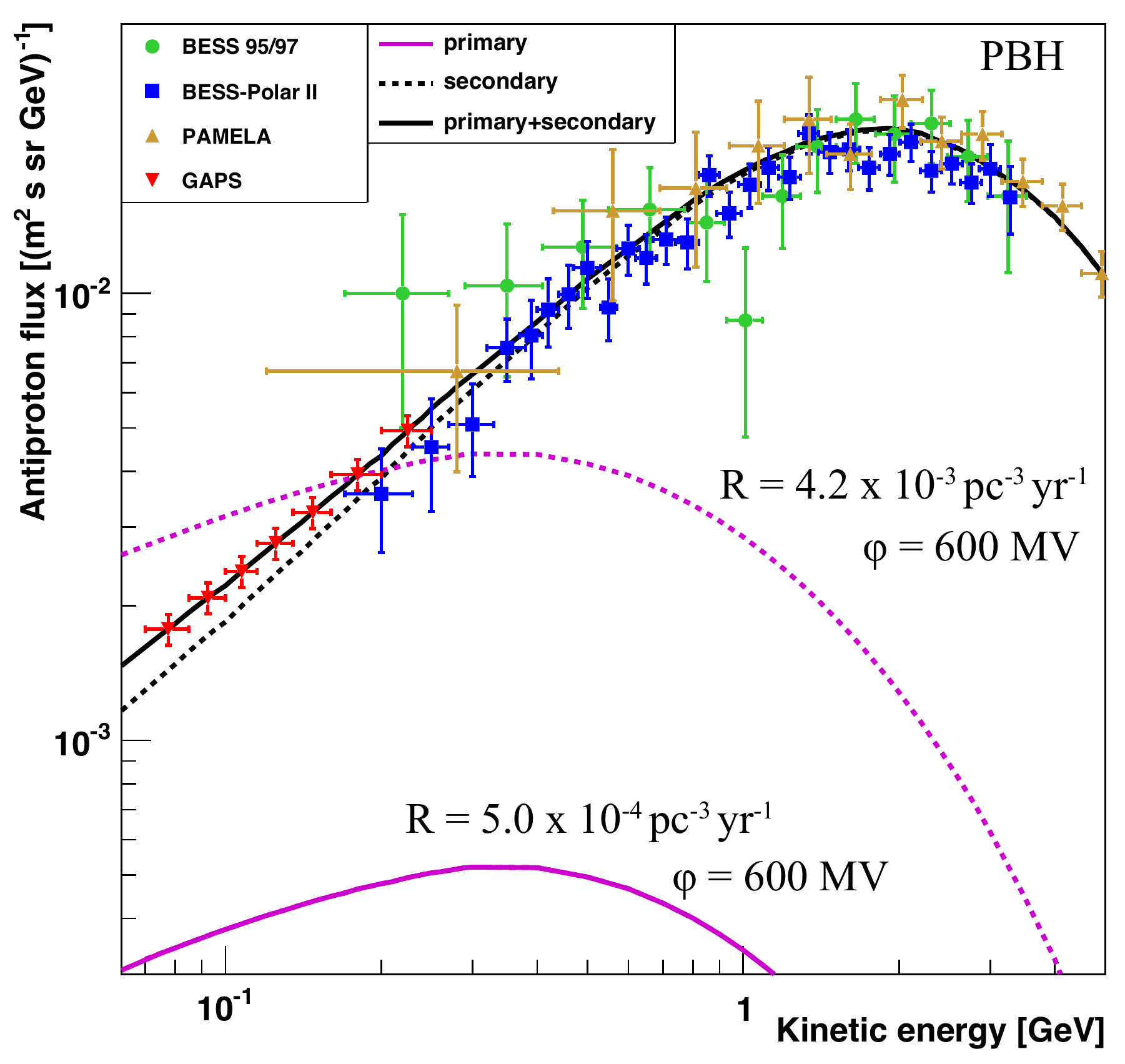}
\end{center}
\caption{The predicted primary antiproton fluxes at the top of the atmosphere (solar-minimum) from evaporating primordial black holes with a local explosion rate of R = 5.0 $\times 10^{-4} $ pc$^{-3} $ yr$^{-1}$ and R = 4.2 $\times 10^{-3} $ pc$^{-3} $ yr$^{-1}$ (purple solid and dashed lines, extrapolated down to $E$ = 65 MeV \cite{Abe2012,Maki1996}), together with the secondary antiproton flux (black dashed line \cite{Abe2012,Mitsui1996}) and the BESS-Polar II \cite{Abe2012}, BESS 95/97 \cite{Orito2000} and PAMELA \cite{Adriani2010} experimental data. The expected GAPS sensitivity for one LDB flight ($\sim$ 40 days, statistical uncertainty only) is also shown for the total antiproton flux (black solid line: primary flux (purple solid line) + secondary flux (black dashed line)).}
\label{PBH}
\end{figure}

Figure \ref{BG} shows two different secondary antiproton flux models (black dashed lines, extrapolated down to $E$ = 65 MeV: BG model 1 for leaky box model with force-field modulation $\phi = 600$ MV \cite{Abe2012,Mitsui1996}, BG model 2 for leaky box drift model with negative solar magnetic field polarity and the tilt angle of the heliospheric current sheet = 15$^{\circ}$ \cite{Abe2012,Bieber1999}), together with the BESS-Polar II \cite{Abe2012}, BESS 95/97 \cite{Orito2000} and PAMELA \cite{Adriani2010} experimental data. The expected GAPS sensitivity for one LDB flight ($\sim$ 40 days, statistical uncertainty only) is also shown for the case of a pure secondary component (BG model 1). The predicted primary antiproton fluxes from dark matter models and evaporating primordial black holes (PBHs, see below) are shown in Figures \ref{neutralino}, \ref{gravitino}, \ref{KK} and \ref{PBH}. 


\begin{figure}[!h]
\begin{center} 
\includegraphics*[width=7.5cm]{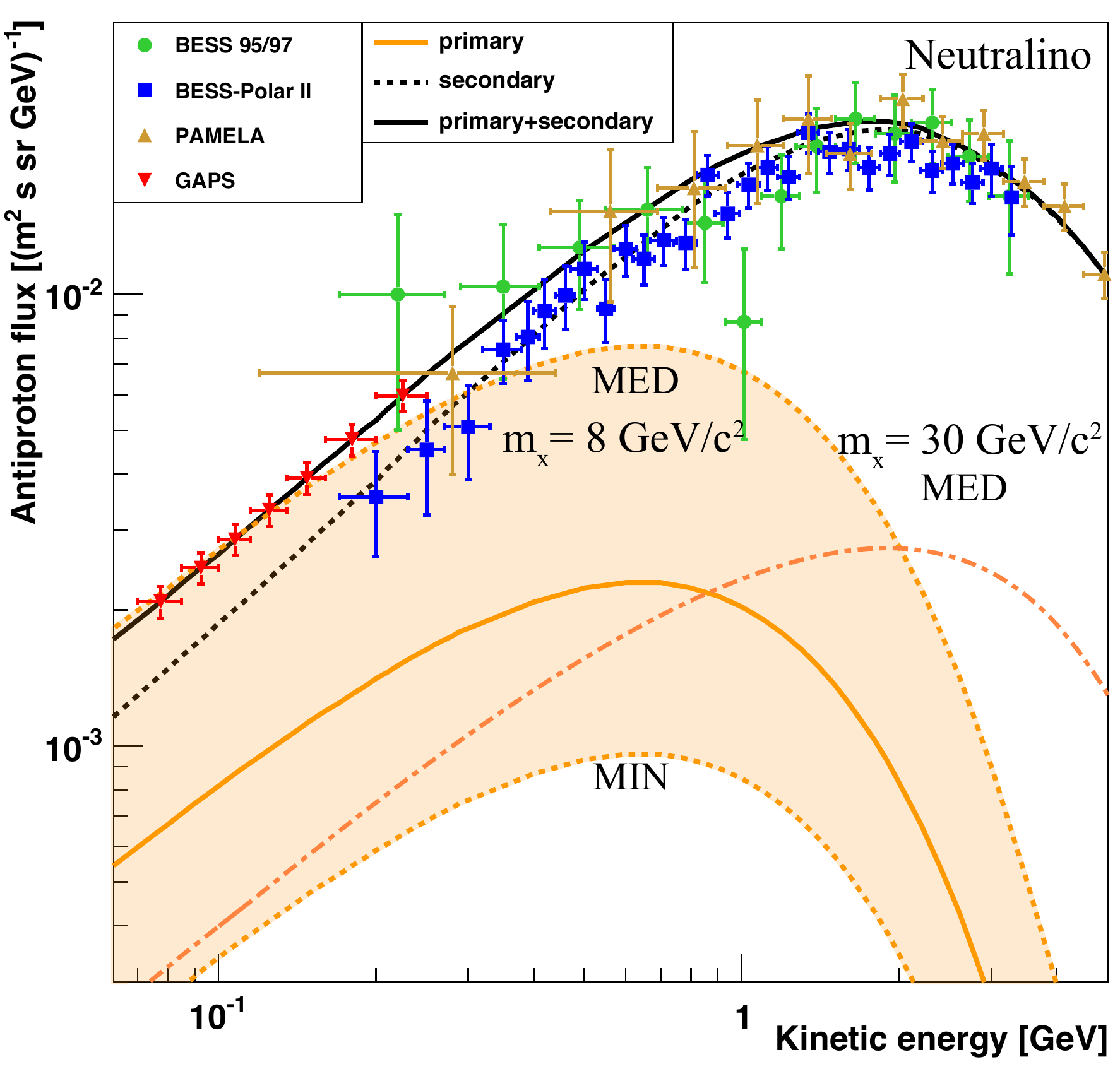}
\end{center}
\caption{The predicted primary antiproton fluxes at the top of the atmosphere from light neutralino dark matter for $m_{\chi} \sim$ 8 GeV/c$^2$ with the medium and minimum astrophysical propagation models (orange dashed lines) and for $m_{\chi} \sim$ 30 GeV/c$^2$ with the medium model (orange dot-dashed line, solar-minimum, extrapolated down to $E$ = 65 MeV \cite{Kappl2012}), together with the secondary antiproton flux (black dashed line \cite{Abe2012,Mitsui1996}) and the BESS-Polar II \cite{Abe2012}, BESS 95/97 \cite{Orito2000} and PAMELA \cite{Adriani2010} experimental data. The expected GAPS sensitivity for one LDB flight ($\sim$ 40 days, statistical uncertainty only) is also shown for the total antiproton flux (black solid line: primary flux (orange solid line) + secondary flux (black dashed line)).}
\label{neutralino}
\end{figure}

The orange dashed lines in Figure \ref{neutralino} represent the primary antiproton flux at the top of the atmosphere (solar-minimum) from light neutralino dark matter for $m_{\chi} \sim$ 8 GeV/c$^2$ with the medium and minimum astrophysical propagation models, while the orange dot-dashed line is for the dark matter mass of $m_{\chi} \sim$ 30 GeV/c$^2$ with medium astrophysical propagation model (extrapolated down to $E$ = 65 MeV \cite{Kappl2012}). The annihilation cross-section used here is the thermal annihilation cross-section ($3 \times 10^{-26}$ cm$^{3}$/s) and the ratio of antiproton flux with the medium propagation model to the one with the minimum propagation model was assumed to be $\sim 8$. While the flux for $m_{\chi} \sim$ 30 GeV/c$^2$ with the medium propagation model is in good agreement with BESS-Polar II data, it is mildly inconsistent with BESS 95/97 data at low energy. On the other hand, although the flux for $m_{\chi} \sim$ 8 GeV/c$^2$ with the medium propagation model disagrees with the BESS-Polar II data, it shows good agreement if the astrophysical propagation model is different from the medium model, if the annihilation cross-section is smaller than the thermal cross-section, or if there is more than one type of dark matter, as discussed above. The orange solid line shows the primary flux for $m_{\chi} \sim$ 8 GeV/c$^2$ with the propagation model that produces three times smaller primary antiproton flux than that produced with the medium model. The expected GAPS sensitivity for one LDB flight ($\sim$ 40 days, statistical uncertainty only) is also shown for the total antiproton flux (black solid line: primary flux (orange solid line) + secondary flux (black dashed line)). 

\begin{figure}[!h]
\begin{center} 
\includegraphics*[width=7.5cm]{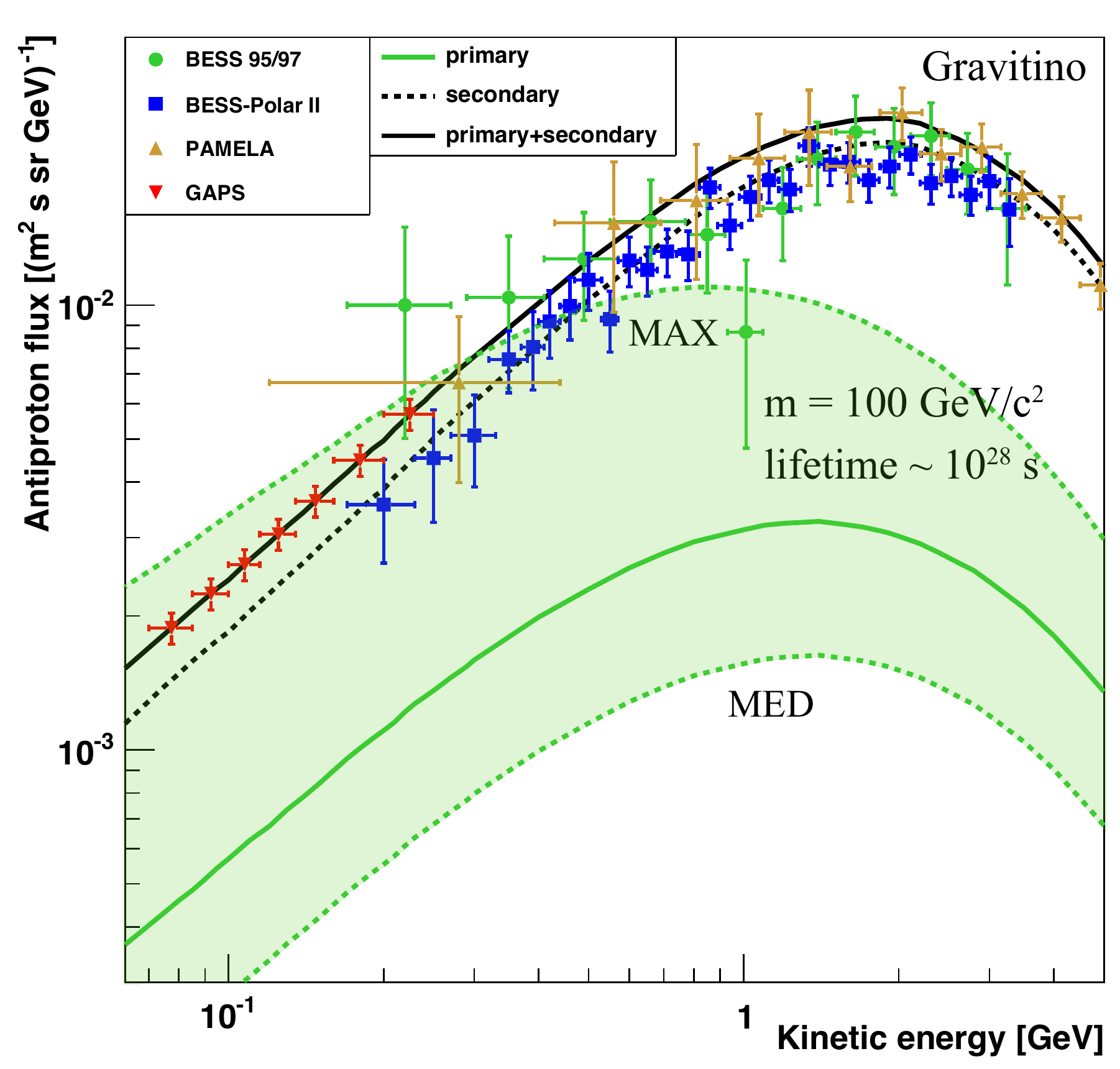}
\end{center}
\caption{The predicted primary antiproton fluxes at the top of the atmosphere (solar-minimum) from gravitino decay for $m_{\chi} \sim$ 100 GeV/c$^2$ with the medium and maximum astrophysical propagation models (green dashed lines), together with the secondary antiproton flux (black dashed line \cite{Abe2012,Mitsui1996}) and the BESS-Polar II \cite{Abe2012}, BESS 95/97 \cite{Orito2000} and PAMELA \cite{Adriani2010} experimental data. The expected GAPS sensitivity for one LDB flight ($\sim$ 40 days, statistical uncertainty only) is also shown for the total antiproton flux (black solid line: primary flux (green solid line) + secondary flux (black dashed line)).}
\label{gravitino}
\end{figure}


Gravitinos, categorized as super-weakly interacting massive particles (Super-WIMPs), are also a dark matter candidate in a supersymmetric model with small R-parity violation \cite{Delahaye2013,Grefe2011}. While the lifetime of the gravitinos can exceed the age of the Universe, considering the suppression of the gravitino decay width by the Planck scale and the small amount of R-parity violation, they can decay into $Z\nu$, $Wl$ and $h\nu$ and produce antiprotons in the final state \cite{Delahaye2013,Grefe2011}. The green dashed lines in Figure \ref{gravitino} indicate the antiproton flux at the top of the atmosphere (solar-minimum, $\phi = 600$ MV) from gravitino dark matter decay for $m_{g}$ = 100 GeV/c$^2$ with the medium and maximum propagation models and decay time $\tau \sim$ 10$^{28}$ s, where the Bino is the next lightest supersymmetric partner (NLSP). The green solid line shows the primary flux with the propagation model that produces two times larger primary antiproton flux than that produced with the medium model. The expected GAPS sensitivity for one LDB flight ($\sim$ 40 days, statistical uncertainty only) is also shown for the total antiproton flux (black solid line: primary flux (green solid line) + secondary flux (black dashed line)). 

The lightest Kaluza-Klein particle can also be a viable dark matter candidate under Z$_3$ symmetry \cite{Lavalle2012,Salati2010,Barrau2005}. The right-handed neutrino can be the lightest Z$_3$ particle (LZP) and stable since it cannot decay into standard model particles. The gauge interaction with the ordinary matter is also suppressed due to the heavy Kaluza-Klein boson, which makes it stable on a Galactic time scale \cite{Lavalle2012,Salati2010,Barrau2005}. The blue dashed lines in Figure \ref{KK} show the antiproton flux at the top of the atmosphere (solar-minimum) from the right-handed neutrino (LZP) with $m_{LZP}$ = 30 GeV/c$^2$ with the medium and maximum propagation models, where the mass of the heavy Kaluza-Klein boson is $m_{KK}$ = 3 TeV/c$^2$ \cite{Lavalle2012,Salati2010,Barrau2005}. The blue solid line shows the primary flux with the propagation model that produces two times larger primary antiproton flux than that produced with the medium model. The expected GAPS sensitivity for one LDB flight ($\sim$ 40 days, statistical uncertainty only) is also shown for the total antiproton flux (black solid line: primary flux (blue solid line) + secondary flux (black dashed line)). 


Aside from dark matter signatures, an excess in the low-energy antiproton flux could also be due to primordial black hole evaporation. Density fluctuations, phase transitions, or the collapse of cosmic strings in the early universe may have formed primordial black holes \cite{Abe2012,Maki1996,Barrau2003,Khlopov2010,Hawking1976}, which could be small enough to have a sufficient rate of evaporation to be observed now, as theoretically predicted by Hawking \cite{Hawking1976}. Antiprotons could be produced as a result of jet-fragmentation in the evaporating primordial black hole. Calculations predict that the primary antiproton flux from evaporating black holes could be as high as the secondary flux at low energy. The purple solid and dashed lines in Figure \ref{PBH} show the primary antiproton fluxes from evaporating primordial black holes with a local explosion rate of R = 5.0 $\times 10^{-4} $ pc$^{-3} $ yr$^{-1}$  and R = 4.2 $\times 10^{-3} $ pc$^{-3} $ yr$^{-1}$ (extrapolated down to $E$ = 65 MeV \cite{Abe2012,Maki1996}). The former satisfies the BESS-Polar II data, while the latter is in agreement with the BESS 95/97 data. The expected GAPS sensitivity for one LDB flight ($\sim$ 40 days, statistical uncertainty only) is also shown for the total antiproton flux (black solid line: primary flux (purple solid line) + secondary flux (black dashed line)). 


\section{Conclusion}

Along with excellent antideuteron sensitivity \cite{Donato2008,Vittino2013,Fornengo2013a}, GAPS has a strong capability to observe signatures of dark matter and evaporating black holes through the measurement of low-energy antiprotons. The antiproton search plays an important role in light dark matter models, as hinted by the recent direct dark matter experiments. While LUX and other dark matter experiments may not be able to completely exclude light dark matter models, especially for an isospin-violating scenario and halo-independent analysis, GAPS can uniquely search for light dark matter with completely different detection methods and backgrounds. 

Since GAPS will be able to detect an order of magnitude more low-energy antiprotons than BESS and PAMELA, dark matter signatures from light neutralinos, as well as gravitinos and Kaluza-Klein particles, could be observed in the low-energy antiproton flux. Evaporating primordial black holes could also be observed through the low-energy antiproton measurement. Even if no excess is found, GAPS will be able to provide strong constraints on models of dark matter, evaporating black holes, and secondary astrophysical flux, as well as the detailed studies of solar modulation, geomagnetic deflection, atmospheric attenuation and atmospheric background production.

\section{Acknowledgments}

This work is supported in the US by NASA APRA Grants (NNX09AC13G, NNX09AC16G) and the UCLA Division of Physical Sciences and in Japan by MEXT grants KAKENHI (22340073). K. Perez's work is supported by the National Science Foundation under Award No. 1202958.





\bibliographystyle{elsarticle-num}

\bibliography{refs}





\end{document}